\documentclass[manuscript]{emulateapj}
\usepackage{natbib}
\usepackage{amsmath}
\usepackage{amssymb}
\DeclareMathSymbol{\circledstar}   {\mathbin}{AMSa}{"46}

\bibpunct{(}{)}{;}{a}{}{,}
\usepackage{txfonts}
\def\tens#1{\ensuremath{\mathsf{#1}}}

\newcommand{\A}           {{\tens{A}}}

\newcommand{\F}           {{\tens{F}}}

\newcommand{\AC}          {{\tens{A_*}}}

\newcommand{\E}           {{\tens{E}}}

\newcommand{\J}           {{\tens{J}}}

\newcommand{\AWP}         {{\it A-Projection}}
\newcommand{\WBAWP}       {{\it WB A-Projection}}
\newcommand{\NBAWP}       {{\it NB A-Projection}}
\newcommand{\MTMFS}       {{\it MT-MFS}}
\newcommand{\Pnu}         {{P_{\nu}}}

\newcommand{\Inu}         {{I_{\nu}}}
\newcommand{\wnt}         {{w_{\nu}^t}}
\newcommand{\ostar}       {{\circledast}}
\newcommand{\dnuno} {{ \left( {\nu-\nu_0}\over{\nu_0} \right)  }}

\def\Adj#1{{adj\left(#1\right)}}
\def\Det#1{{det\left(#1\right)}}

\begin{document}
%
\title{Wide-field wide-band interferometric imaging:
  The WB~A-Projection and hybrid algorithms} 


\shorttitle{The WB A-Projection algorithm} 
\shortauthors{Bhatnagar, Rau \& Golap}
\author{S.~Bhatnagar}
\affil{National Radio Astronomy Observatory, Socorro, NM - 87801, U.S.A.}
\email{sbhatnag@nrao.edu}
\author{U.~Rau}
\affil{National Radio Astronomy Observatory, Socorro, NM - 87801, U.S.A.}
\email{rurvashi@nrao.edu}
\author{K.~Golap}
\affil{National Radio Astronomy Observatory, Socorro, NM - 87801, U.S.A.}
\email{kgolap@nrao.edu}

\thanks{The National Radio Astronomy Observatory is a facility of the
  National Science Foundation operated under cooperative agreement by
  Associated Universities, Inc.}

\date{Received: 08-08-2012;  Accepted: 04-17-2013}

\begin{abstract}

  Variations of the antenna primary beam (PB) pattern as a function of
  time, frequency and polarization form one of the dominant
  direction-dependent effects at most radio frequency bands.  These
  gains may also vary from antenna to antenna.  The \AWP\ algorithm,
  published earlier, accounts for the effects of the narrow-band
  antenna PB in full polarization.  In this paper we present the
  Wide-Band A-Projection algorithm (\WBAWP) to include the effects of
  wide bandwidth in the A-term itself and show that the resulting
  algorithm simultaneously corrects for the time, frequency and
  polarization dependence of the PB.  We discuss the combination
  of the \WBAWP\ and the Multi-term Multi Frequency Synthesis (\MTMFS)
  algorithm for simultaneous mapping of the sky brightness distribution
  and the spectral index distribution across a wide field of view.  We
  also discuss the use of the
  narrow-band \AWP\ algorithm in hybrid imaging schemes that account for
  the frequency dependence of the PB in the image domain.
\end{abstract}  


\maketitle
%

\section{Introduction}

Observations in the radio band offer distinct, and
often times unique, scientific advantages in probing certain areas of
astrophysical research (e.g in the detection of the EoR signal,
studies of the high-redshift universe in general, large-scale
structure formation, early galaxies, etc.).

All next generation radio telescopes, many in operation now, offer at
least an order of magnitude improvement in the sensitivity and angular
resolution compared to the telescopes operated in the past decades.
The two key instrumental parameters which afford such high
sensitivities, impact the imaging performance and are significantly
different from previous generation telescopes are: 1) the wide
instantaneous fractional bandwidths, and 2) larger collecting area.
The effects of wide instantaneous fractional bandwidths that classical
calibration and imaging algorithms ignore, lead to errors higher than
the sensitivity that these new telescopes offer.  Examples, relevant
for some of the telescopes already in operation include the effects of
time and frequency variant primary beams, frequency dependence of the
emission from the sky and antenna pointing errors.  The effects of
wide fractional bandwidth and ionospheric phase screen limit the
imaging performance below $\sim$1GHz.  Additionally, significant
variations in the shape of the wide-band primary beams (PB) for
aperture array telescopes leads to errors of similar magnitude.  All
these effects form the general class of problems referred to in the
literature as ``direction dependent effects'' or DD effects.

Both, wide fractional bandwidths and larger collecting area lead
to many orders of magnitude increase in the data volume, putting
severe constraints on the run-time performance of the algorithms for
calibration and imaging.  Furthermore, the cost of software
development and maintenance also scales with algorithm complexity.
Efficient algorithms to simultaneously account for all time-,
frequency- and polarization-dependent DD effects which can also
process large data volumes without significantly increasing
algorithmic and software complexity are required.

In the following sections we discuss various possible approaches
to full-beam wide-band continuum imaging.  We present a modification of the
\AWP\ algorithm which we call the Wide-band \AWP, or \WBAWP\ algorithm where
a modified A-term also compensates, to a large extent, the frequency
dependence of the PB.  
We also discuss the use of the unmodified \AWP\ algorithm
\citep[henceforth referred to as Paper-I]{AWProjection}, which we call
the Narrow-band \AWP, or NB \AWP, along with various forms of
image-plane normalizations for wide-band continuum imaging and the
resulting issues and limitations.

\section{Theory}
\label{SEC:THEORY}
Using the notation developed by \cite{HBS1}, full polarimetric
measurements from a single baseline calibrated for the effects of
direction-independent gains, can be described by the following
Measurement Equation
\begin{equation}
\label{EQ:ME}
\vec{V}^{Obs}_{ij}(\nu,t) = W_{ij}(\nu,t) \int P^{Sky}_{ij}(\vec{s},\nu,t)
\vec{I}(\vec{s},\nu) e^{\iota \vec{b}_{ij} \cdot \vec{s}}d\vec{s}
\end{equation}
where $\vec{V}^{Obs}_{ij}$ are the observed visibility samples
measured by the pair of antennas designated by the subscript $i$ and
$j$, separated by the vector $\vec{b}_{ij}$ and weighted by the
measurement weights $W_{ij}$.  $P^{Sky}_{ij}$ is the radio-Mu\"eller
matrix\footnote{This matrix as used in radio interferometric
  literature differs from that used in the optical literature only in
  that in radio it is written in the polarization basis (circular or
  linear polarization) while in the optical literature it is written
  in the Stokes basis.  These radio and optical representations are
  related via a Unitary transform \citep{HBS1}.} in the image domain
representing the full polarization description of the antenna primary
beams as a function of the direction $\vec{s}$, frequency $\nu$ and
time $t$ and $\vec{I}$ is the image vector.  The vectors $\vec{V}$ and
$\vec{I}$ are full polarization vectors in the data and image domain
respectively.  $P^{Sky}_{ij}$ and $\vec{I}$ are the unknowns in this
equation.

Equation~\ref{EQ:ME} cannot be directly inverted as, in general, it is
not a Fourier transform relation.  It is also sampled only at a
limited number of points, and therefore the data has insufficient
information to allow an exact solution.  Estimation of $\vec{I}$ is
therefore typically done via iterative non-linear
$\chi^2$-minimization \citep{CORNWELL_AIPS++_2, IMAGING_THEORY_IEEE}.
Below we briefly review the theory of imaging with \AWP\ to correct
for the time and polarization dependence of $P^{Sky}_{ij}$ in
narrow-band imaging and motivate the need for a Wide-band \AWP\
algorithm to also correct for frequency dependence of $P^{Sky}_{ij}$
in wide-band imaging.

\subsection{Imaging with \AWP}
\label{SEC:THEORY_IMAGING}

To clarify the full-polarization nature of the \AWP\ algorithm, we
define the outer-convolution operator and denote it by the symbol
$\ostar$.  The outer-convolution operator is similar to the
outer-product operation used in the direction-independent (DI)
description of \cite{HBS1} with a minor difference.  The
element-by-element algebra of the outer-convolution operator is the
same as that of the outer-product operator, except that the complex
multiplications in outer-product are replaced by convolutions.  Using
the outer-convolution operator and the sub-scripts $i$ and $j$ to
explicitly denote the antenna pair for baseline $i-j$, the A-matrix
used in \AWP\ at a frequency $\nu$ and time $t$ can be written in
terms of antenna based quantities as
\begin{equation}
\label{EQ:A}
\A_{ij} = \J_i \ostar {\J}^{*}_j
\end{equation}
where
\begin{equation}
\J_i = 
\left[ \begin{array}{ll}
   {\E}^p_i  & {\E}_i^{p\rightarrow q} \\
   {\E}_i^{q\rightarrow p} & {\E}^{q}_i \\
 \end{array} \right]
\end{equation}
$\E^p$ and $\E^q$ are the polarized antenna aperture illumination
patterns for the two polarization states.  The off-diagonal terms are
the leakage patterns.  $\A_k$ is the DD equivalent of the $4\times 4$
DI Mu\"eller matrix for a given antenna pair.  The elements of
$\A_{ij}$ are the complex convolution of the two antenna aperture
illumination patterns $\left({\E}^p_i \star {\E}^{p^*}_j\right)$,
$\left({\E}^p_i \star {\E}^{{p\rightarrow q}^*}_j\right)$, etc.  For
comparison, the elements of $\J_i \otimes {\J}^{*}_j$ would be
$\left({\E}^p_i \cdot {\E}^{p^*}_j\right)$, $\left({\E}^p_i \cdot
  {\E}^{{p\rightarrow q}^*}_j\right)$, etc.

To keep the notation simple, in the following description we use a
single sub-script $k\equiv (ij,\nu,t)$ to refer to a measurement from
a single baseline $ij$, at a spot frequency $\nu$ and an instant in
time $t$.  The vectors $\vec{V}$ and $\delta$ are full polarization
vectors whose elements are 2D functions in the visibility plane (the
uv-plane).  Elements of $\vec{V}$ are the 2D visibility data and
elements of $\delta$ are 2D Delta functions representing the
uv-sampling function for the data sample $k$.  The super-scripts
$obs$, $M$ and $\circ$ refer to the observed, model and true values
respectively.  

Using the notation described above, the $\chi^2$ can be written as
\begin{equation}
\chi^2 =\sum_k \vec{V}^{R^\dag}_k \Lambda_k \vec{V}^R_k
\end{equation}
where $\vec{V}^R_k = \vec{V}^{Obs}_k - \vec{V}^M_k$ and $\Lambda_k$ is
inverse of the noise covariance matrix.  The vector
$\vec{V^{Obs}}$ can be expressed in terms of $\A$ as
\begin{equation}
\label{EQ:DDME}
\vec{V}^{Obs}_k = \left(\A^\circ_k \star \vec{V}^\circ \right) \delta_k
\end{equation}
Note that, as mentioned before, the elements of ${\A^\circ}$,
$\vec{V}^\circ$ and $\delta_k$ are 2D functions.  The symbol '$\star$'
represents the element-by-element convolution.  $\vec{V}^\circ$ --
without a sub-script -- represents the {\it true} continuous Coherence
function.  $\vec{V}^{obs}_k$ represents a sample of this Coherence
function measured at the parameters represented by sub-script~$k$.

The calibration matrix for Eq.~\ref{EQ:DDME} to correct for the
effects of ${\A}^\circ_k$ is ${\A}^{\circ^{-1}}_k$ given by
\begin{equation}
\label{EQ:AINV}
\A^{\circ^{-1}}_k=\frac{\Adj{{\A}^\circ_k}}{\Det{{\A}^\circ_k}}  
\end{equation}
The equivalence between Eq.~\ref{EQ:AINV} as a generalized
direction-dependent (DD) calibration and
standard direction-independent (DI) calibration 
is discussed in more detail in section~\ref{SEC:GENERALIZEDCALIB}.

As in DI calibration where calibration is done by the application
of the inverse of the appropriate Mu\"eller matrix, correction for the
effects of $\A$ requires the application of $\A^{-1}$.  The difference
between DI and DD calibration is that while the operator for the
application of the DI calibration matrix to the data is the matrix
multiplication operator, for DD calibration this operator is the
element-by-element convolution operator (the $\star$ operator in
Eq.~\ref{EQ:DDME}).

Since DD calibration fundamentally cannot be separated from imaging,
the application of the $\A^{-1}$ matrix is done via the \AWP\
algorithm.  This is achieved in two steps.  The term in the numerator
of Eq.~\ref{EQ:AINV}, $\Adj{{\A}_k}$, is applied during re-sampling of
the observed data (the right hand side of Eq.~\ref{EQ:DDME}) on a
regular grid using convolutional gridding with 
${\A}^{M^\dag}_k$, a model of $\A^{\circ^{\dag}}_k$, as the convolution function.  The
resulting gridded data is accumulated in the data domain and then
Fourier transformed to compute the continuum image.  The
scaling by the denominator of Eq.~\ref{EQ:AINV} is done by also
accumulating $\A^{M}_k$ and diving the image by its Fourier transform.
The resulting image using \AWP\ is given by
\begin{equation}
\label{EQ:AWPIM}
I^R  =  \frac{\F \sum_k \Adj{\A^{M}_k} \star \left(\A^\circ_k \star
    V^\circ \right) \delta_k}{det\left(\F \left[\sum_k {\A}^M_k\right]\right)}
\end{equation}
This effectively applies the DD calibration operator ${\A}^{-1}$ and
corrects for its effects, {\it provided} ${\A}^M$ is a close enough
approximation of ${\A}^\circ$.  Further details, results and discussion
on the imaging performance of \AWP\ are in Paper-I.

For continuum imaging, the accumulation for all $k$ in
Eq.~\ref{EQ:AWPIM} can be done in either domain (data or image
domain).  Continuum imaging of wide-band data using the MFS approach
is done by accumulation in the data domain.  Since the \AWP\ algorithm
does not explicitly account for the frequency dependence of $\A$, an
algorithm to project-out this frequency dependence {\it before}
accumulation is required.  The \WBAWP\ algorithm for this is described
in section~\ref{SEC:WBAWP_ALGO}.  Various hybrid imaging algorithms using
the NB-\AWP\ algorithm and accumulation in the image domain are also
possible.  These are discussed in section~\ref{Sec:NBforMFS}.

\subsubsection{Algorithmic steps for \AWP}

For completeness and as a reference for later discussions, the
algorithmic steps for MFS imaging using the \AWP\ algorithm described
in Paper-I are repeated below:
\begin{enumerate}
\item Initialize the model and the residual images $I^M$ and
  $I^R$

\item {\bf Major cycle:}
\label{STEP:1}
\begin{itemize}
\item
  Predict the model data and accumulate $\A_k$ as:
  \begin{eqnarray}
    \vec{V}^{M}_{k}  &=& \sum_k \Adj{\A_{k}}\star \left(\F^{-1}\vec{I}^{M}\right)\nonumber\\
    \overline{\A^M} &=&  \sum_k {\A}^M_k \nonumber
  \end{eqnarray}

\item Compute the residual data $\vec{V}^{R}_{k}=\vec{V}^{Obs}_{k} - \vec{V}^{M}_{k}$

\item Use Eq.~\ref{EQ:AWPIM} to compute the continuum residual image
  as $\vec{I}^R = \left( \F \sum_k \vec{V}^R_k \right)/ \Det{\F \overline{\A^M}}$

\end{itemize}
\item {\bf Minor cycle:} Invoke the appropriate minor-cycle algorithm
  using $\vec{I}^R$ to solve for image-plane parameters and update the
  model image $\vec{I}^M$.

\item If not converged, go to Step~\ref{STEP:1}.
\end{enumerate}

Since $\A_k$ does not change from one major cycle to another,
accumulation of $\overline{\A^M}$ is done in the first major cycle and
cached for use in subsequent major cycles.

\subsubsection{A-Projection: A direction-dependent gain correction
  algorithm}
\label{SEC:GENERALIZEDCALIB}
The antenna illumination pattern is essentially a direction-dependent
description of antenna based complex gains in the data domain.  $\A_k$
is a direction-dependent generalization of the $G$-Jones matrix -- the
{\it direction-independent} Mu\"eller matrix for antenna gains in the 
\cite{HBS1} formulation and since the \AWP\ algorithm corrects for the
effects of $\A_k$, it can be thought of as an algorithm for DD
calibration.

To establish the equivalence between DD corrections via \AWP\ and DI
antenna-based complex gain correction, we note that Eq.~\ref{EQ:DDME}
is the DD equivalent of DI measurement equation given by:
\begin{equation}
\label{EQ:DIME} 
\vec{V}^{obs}_{ij} = G_{ij} \cdot \vec{V}^\circ_{ij}
\end{equation}
${\A}_k$ in Eqs.~\ref{EQ:A} and \ref{EQ:DDME} is the DD equivalent of
$G_{ij}$ and the outer-convolution ('$\ostar$') and the '$\star$'
operators are the DD equivalent of the outer-product ('$\otimes$') and
element multiplication.  Calibration for $G_{ij}$ is done by
multiplying Eq.~\ref{EQ:DIME} by $G^{-1}_{ij}$ given by
\begin{equation}
\label{EQ:GINV}
G^{-1}_{ij}=\frac{G_{ij}^{*}}{\left|G_{ij}\right|^2}  
\end{equation}
For calibration, Eq.~\ref{EQ:AINV} is the DD equivalent of the above
equation for the DD calibration.  

For an intuitive understanding, we note that for the simpler case where
the off-diagonal elements of $\J$ are negligible, $\Adj{{\A}_k} =
{\A}^\dag$ and
$\Det{{\A}_k}=trace\left(\A_k\right)$=${\E}^{pp}_k{\E}^{qq^*}_k$.  For
this simpler case, examination of Eqs.~\ref{EQ:AINV} and \ref{EQ:GINV}
shows the equivalence between DI gain calibration and the \AWP\
algorithm more clearly.  The process of imaging using the $\Adj{\A_k}$
and normalizing the resulting image by $\Det{\sum_k \A_k}$ therefore,
even for the general case, is the DD generalization of the
antenna-based complex gain calibration.

\begin{figure*}[ht]
\centering
 \includegraphics[width=8cm]{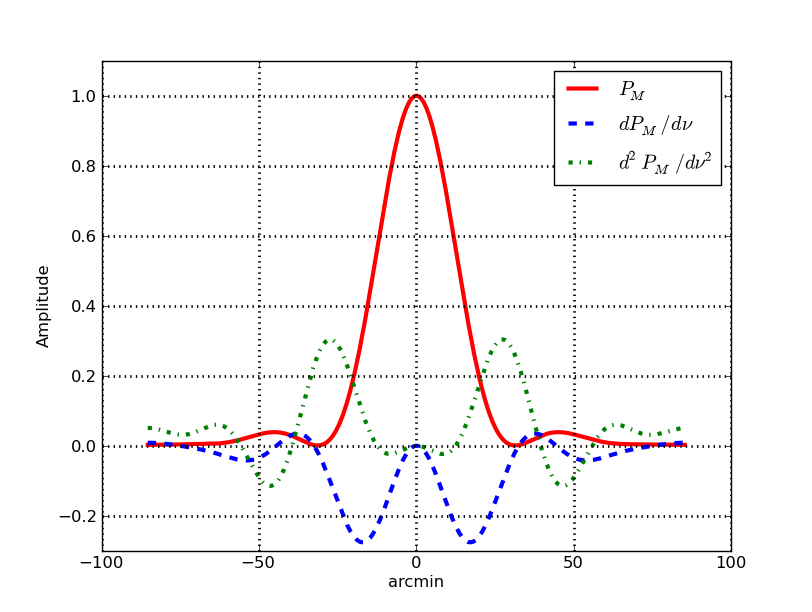}
 \includegraphics[width=8cm]{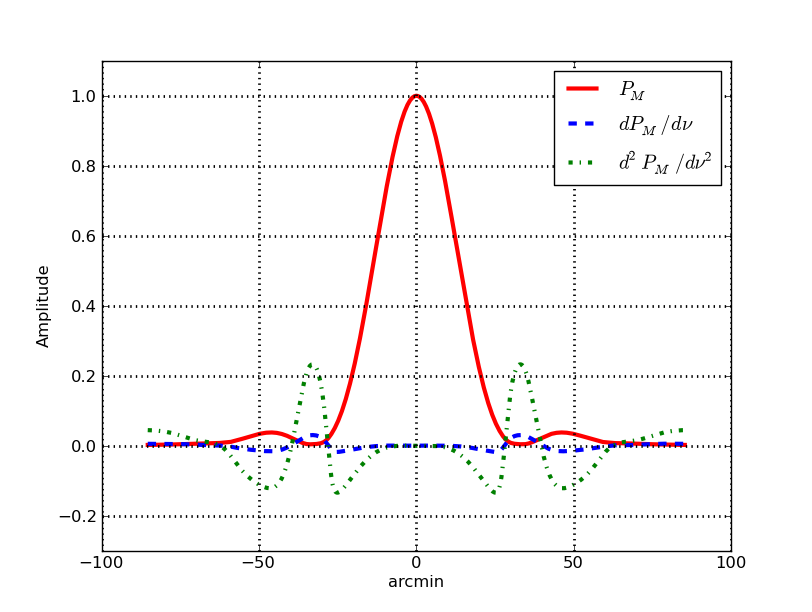}
 \caption{The plot in the left panels shows the one dimensional cuts
   through the PB model at a reference frequency (continuous red line)
   for the VLA and its first (dashed blue line) and second derivatives
   (dash-dot green line) with respect to frequency.  The plot on the
   right shows the same cuts through the $P_{eff}(\nu)$ -- the
   effective frequency-independent PB.}
\label{FIG:WBPB_COEFF_CUT}
\end{figure*}
\section{The WB A-Projection algorithm}
\label{SEC:WBAWP_ALGO}

$\Adj{\A_k}$, when used in the \AWP\ algorithm, corrects for the
polarization and other DD effects that can be encoded in the phase of
$\A^\circ$ (e.g. time-varying gains due to polarization squint,
ionospheric phase-screen, etc.).  It
is however not a conjugate operator for variations along the time or
frequency axis.

The image domain effects of the time varying gains are largely in the
amplitude scaling only (i.e., they do not disperse the flux in the
image domain).  Since the minor cycle algorithms typically assume that
the sky brightness distribution is {\it time-invariant} and do not
parametrize the model image in time, the effects of such variations
can be ignored in the transform from data to image domain.  If the
deviations from the average value are small (e.g. for antenna
arrays and long integrations where time variability is cyclic, or
antennas with three-axis mounts where beams do not rotate on 
the sky) the
model prediction stage, which properly includes these effects,
corrects for time variability in an the iterative deconvolution
scheme.

Frequency dependence of $\A^\circ$ also varies with time (and
direction).  This variation is not cyclic and its maximum deviation
from the average value increases with fractional bandwidth.  Due to
this, as for time varying gains, while the frequency dependence of
$\A^\circ$ in the inner part of the main lobe of the PB can also be
corrected via the model prediction stage, the convergence is
significantly slower (requiring more major cycles and hence higher
computing).  Alternatively, this frequency dependence can be absorbed,
to some extent, in the multi-term MFS (MT-MFS) minor cycle algorithm
which solves for the time-invariant frequency dependence in the image
plane.  But this requires more Taylor-terms, which also is inadvisable
(see section~\ref{SEC:METHOD_2}).

Correction for the time-variable frequency dependent effects of
$\A^\circ$ requires a wide-band version of the $\A^M$ matrix such that
$\Adj{{\A}^M_k} \star {\A}^\circ_k$ results in a function which does
not vary with frequency (i.e., $\Adj{\A^M}$ is also a conjugate
operator for frequency).  The frequency dependence will then be
projected-out prior to accumulation, resulting in an image corrected
for the frequency dependent effects of $\A$.

\subsection{The wide-band $\A$ operator}
\label{SEC:ATERM}

For the reasons givens above, as well as to keep the parameters for
modeling the instrumental effects in the data domain separate from parameters
for modeling the sky brightness distribution, we need to construct the
$\AC(\nu)$ matrix which projects-out the dependence on frequency
during imaging.  One such possibility is the following:
\begin{equation}
\label{EQ:CONJ-PB}
\AC(\nu) = \F^{-1} \left[ \frac{P_{eff}(\nu) P^*_{eff}(\nu)}{P(\nu)}\right] 
\end{equation}
where $P(\nu) = \F^{\dag} \A(\nu)$ and $P_{eff}(\nu)$ is the desired
effective PB which varies minimally with frequency.
In the data domain, $\Adj{\A_*(\nu)}\star\A(\nu)$ can be
shown to be frequency-independent to high orders, and this use of $\AC(\nu)$
as a model for $\A(\nu)$ in the reverse transform can correct for the
frequency dependence of $\A(\nu)$.  However it also has a large
support size, and therefore, in itself, is not an efficient reverse
transform operator.

\subsection{The  Conjugate frequency}

Equation~\ref{EQ:CONJ-PB} is valid for any frequency dependence.  For
the special case of PB scaling with frequency, we explore usable
approximations for $\AC$, we define the {\it conjugate frequency}
$\nu_*$, given by\footnote{This expression is arrived at by using a
  gaussian approximation for the PB and imposing the condition that
  $P(\nu_*) P(\nu) = P_{eff}(\nu)$}:
\begin{equation}
\label{EQ:CONJ-NU}
\nu_* = \sqrt{2\nu_{ref}^2 - \nu^2 }
\end{equation}
where $\nu_{ref}$ is the reference frequency of the continuum image,
and examine the effects of choosing $\A_*(\nu) \equiv \A(\nu_*)$.

Using the same model for $\A$ as used in Paper-I (i.e. a model for the
VLA antenna PB), one dimensional cuts through the model PB given by
$P_M(\nu)=\F A(\nu)$, the effective PB given by $P_{eff}(\nu)=\F
\left[A(\nu_*)\star A(\nu)\right]/\left[\F A(\nu_{ref})\right]$ and
the first and second derivatives of $P_{eff}(\nu)$ with respect to
frequency are shown in Fig.~\ref{FIG:WBPB_COEFF_CUT}.  A comparison of
the derivatives of $P(\nu)$ and $P_{eff}(\nu)$ with frequency, shown
in the two panels as blue dashed lines, shows that the effective PB is
frequency-independent to the first order.  While it changes in
structure, the maximum second derivative remains almost the same in
magnitude.  These figures show that the approximation in
Eq.~\ref{EQ:CONJ-NU} and use of $\A(\nu_*)$ is good enough for imaging
data which are not sensitive to the higher order frequency dependent
effects.  This approximation is useful since it can be easily
implemented, is appropriate for the sensitivity of current telescopes
and covers a large fraction of scientific observations for
simultaneous Stokes-I and spectral index mapping.  The frequency
dependence in $\A(\nu)$ is reduced overall by an order of magnitude.
When used in the \AWP\ algorithmic steps in
section~\ref{SEC:GENERALIZEDCALIB}, it effectively corrects for the
frequency dependence of the PB {\it prior} to the accumulation along
the frequency axis in Eq.~\ref{EQ:AWPIM}.  For future, more sensitive
telescopes which will be sensitive to second order frequency dependent
effects also, $\AC$ as in Eq.~\ref{EQ:CONJ-PB} may be required.
However, when software implementation itself requires partitioning
along the time and/or frequency axis, hybrid approaches discussed in
section~\ref{Sec:NBforMFS} may be more efficient.

In practice, there might be multiple terms that make up the A-term,
some of which may scale with frequency while others may not.
E.g. Aperture Illumination will scale with frequency, but
pointing-offset term or resonances effects will not.  The terms that
do not scale with frequency can be applied as multiplicative terms to
$\A(\nu_*)$ during gridding.  We have tested this approach to work for
parallel hand polarization effects.  While we have not tested it, we
think this may also work for cross-hand polarizations.

To avoid confusion and for brevity, we will refer to the algorithm in
Paper-I as the Narrow-band \AWP\, or \NBAWP\ algorithm, and refer to
the use of $\A(\nu_*)$ (instead of $\A(\nu)$) for the data-to-image
domain transform as the \WBAWP\ algorithm.

\begin{figure*}[ht!]
\centering
 \includegraphics[width=6.5in]{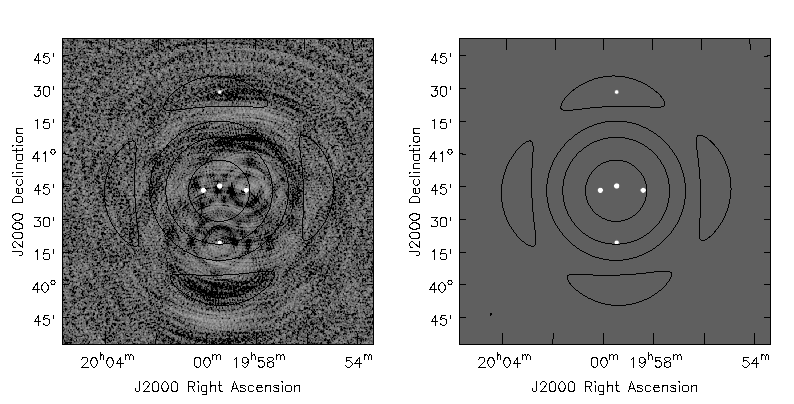}
 \caption{This figure shows imaging performance before and after applying
   corrections for the time and frequency dependence of the PB during imaging.
   The sky is assumed to have a flat-spectrum, and standard MFS imaging is done.
   Both restored images are shown at the same gray-scale, stretched to emphasize 
   artifacts.    Contours are drawn at the 0.02, 0.1 and 0.5 (HPBW) levels of the
   time-and-frequency averaged Primary Beam.
   No noise was added to the simulated visibilities, in order to clearly illustrate the
   noise-like artifacts produced by time-variable DD-effects.
   \newline
   LEFT : Standard MFS-imaging and deconvolution, using a prolate-spheroidal gridding
   convolution function. Dominant errors are due to the time and frequency 
   variability of the PB.
   Off-source RMS  : $4\times 10^{-4} Jy$, Peak Residual : $1.8\times 10^{-3} Jy$
   \newline
   RIGHT : MFS-imaging and deconvolution, using \WBAWP\ to account for
   both time and frequency variability during gridding. 
   Off-source RMS  : $1.5\times 10^{-7} Jy$, Peak Residual : $7\times 10^{-7} Jy$
}
\label{FIG:RESULTS_SIM_FLAT}
\end{figure*}

\subsection{Algorithm Validation}
\label{SEC:RESULTS}

The image deconvolution algorithm described in
section~\ref{SEC:WBAWP_ALGO} was tested using simulated wide-band data
with 66\% fractional bandwidth.  The VLA C-array was used for antenna
configuration and the observations covered Hour Angle range of
$\pm3^h$.  The model for the PB used in Paper-I was scaled by
frequency and rotated with Parallactic Angle to simulate time-varying
frequency dependent effects.  To clearly highlight the effects of time
and frequency dependence of $\A$, we used a model of the sky
consisting of five point sources located at 0.99, 0.83, 0.60 and 0.11
levels of the PB within the main lobe and one source located in the
first side lobe (PB gain of 0.025).  All the point sources were
assigned a flux of 1 Jy with flat spectra.  The effective
spectral-indices due to the primary-beam at the five locations are
-0.026,-0.38, -1.0,-5.32 and +0.47 respectively.  No noise was added
to these simulations, and all imaging and deconvolution runs were with
a loop-gain of 0.2.

Figure~\ref{FIG:RESULTS_SIM_FLAT} shows deconvolved images produced
without (first panel) and with (second panel) \WBAWP\ gridding. This
comparison demonstrates that with an accurate model of the Primary
Beam, it is possible to correct-for its {\it time- and
  frequency-variability} down to numerical precision levels in
wide-band wide-field imaging.

\section{Applying NB A-Projection for wide-band imaging}
\label{Sec:NBforMFS}
%

\NBAWP\ was
designed for a single reference frequency and does not automatically
account for the frequency-dependence of P during gridding.
However, it can still be used for wide-band
imaging, as long as the frequency dependence of the far-field pattern
is known and characterized by $P_{\nu}$.  Several algorithmic options
exist, all with different numerical approximations and computing load.
This flexibility allows the implementation to be tuned according to
the available computing resources, architecture of the hardware
platform and the desired imaging accuracy.

\subsection{Cube imaging + Cube deconvolution}
The simplest approach is spectral-cube imaging where each frequency channel 
(with its limited $uv$ coverage) is treated separately.
\NBAWP\ is applied as is per channel, 
and the minor cycle run independently per channel. 
A continuum image is later constructed
by adding together deconvolved and restored images from all channels.

The residual image per channel can be approximately 
re-written from Eq.~\ref{EQ:AWPIM} in the image domain, 
for the case where the aperture illumination functions are identical for
all baselines and times, and only one polarization-pair is being imaged.
\begin{equation}
I^{R}_{\nu} = {{ {\Pnu \cdot \left( I^{psf}_{\nu} \star \left(\Pnu \cdot I^{sky}_{\nu}\right) \right)}} \over {\Pnu^2} }
\label{Eq:imnarrow}
\end{equation}
where $I^{psf}_{\nu} = F^{-1}\sum_k \delta_{k,\nu}$ ($k\equiv ij,t$) is the point spread function for one channel
and $\Pnu = F^{-1} A_{\nu}$.
In this expression, the division by $\Pnu^2$ implies a {\it flat-sky}\footnote{
Flat Sky normalization : $P^2$ in the denominator of Eq..\ref{Eq:imnarrow} 
gives a residual image in which the peak brightness 
is free from the primary beam but the noise level is position dependent. 
$I^{R}$ does not strictly follow a convolution equation, and may require 
shallow minor cycles if the PSF is not well behaved.
The output model image from the minor cycle 
represents only the true sky $I^{sky}$.
} normalization, but a {\it flat-noise}\footnote{
Flat Noise normalization : $P$ in the denominator of Eq..\ref{Eq:imnarrow} 
instead of $P^2$ gives a residual image representing the signal-to-noise ratio
at all pixels. Also, $I^R$ satisfies a convolution equation, 
allowing for deeper deconvolution in the minor cycles. 
The model image will however represent $P \cdot I^{sky}$, and a post-deconvolution
division of this model by $P$ will be required.  
} normalization may be used instead.

This method is straightforward and will suffice for modest 
imaging dynamic ranges and uncomplicated spatial structure (point sources).  
However, the angular resolution of the continuum image and any
estimate of the sky spectrum will be limited to that of the 
lowest frequency in the band. Also, reconstruction uncertainties may
be inconsistent across frequency when there is insufficient $uv$-coverage
per channel or complicated spatial structure, leading to spurious spectral
structure. 

In this paper, we are focusing on imaging problems that require more accuracy and
dynamic-range than what the above offers. In the next two sections, we 
discuss \AWP\ in the context of multi-frequency-synthesis (MFS)
where the combined $uv$-coverage is used for model reconstruction (minor cycle).

\begin{figure*}[ht!]
  \centering
  \includegraphics[width=6.9in]{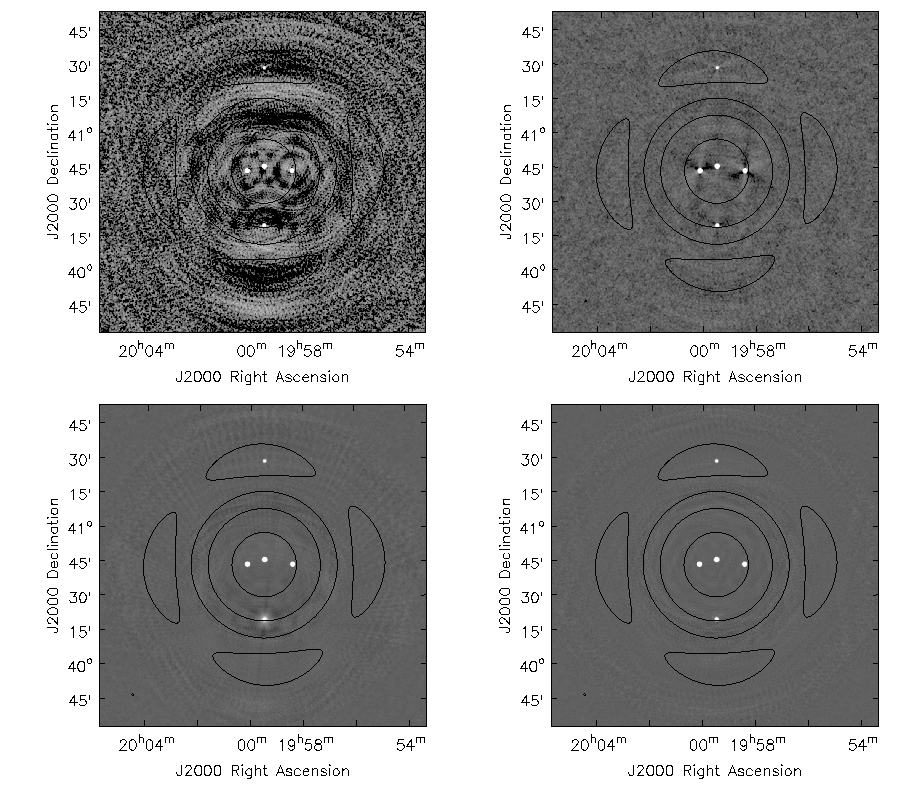}

  \caption[]{These figures compare the imaging performance before and after applying
    corrections for the time and frequency dependence of the PB during imaging.
    All restored images are shown at the same gray-scale, stretched to emphasize 
    artifacts. Contours are drawn at the 0.02, 0.1 and 0.5 (HPBW) levels of the
    time-and-frequency averaged Primary Beam.   Results from four algorithms 
    described in Sec.~\ref{SEC:RESULTS_MTMFS} are compared
    here (MFS+SI, \MTMFS+SI, \MTMFS+\AWP, \MTMFS+\WBAWP). 
    RMS and peak-residuals are listed in the table below.
    \newline
    \newline
    \begin{tabular}{|p{0.5in}|p{0.8in}|p{1.2in}|p{0.5in}|p{0.8in}|p{2.1in}|}
      \hline
      Panel & Algorithm & Description & RMS & Peak Residual & Comments\\
      & & & (Jy/beam) & (Jy/beam) & \\
      \hline
      Top Left & MFS + SI & Standard Wide-band Imaging &$6\times 10^{-4}$ & $2.3\times10^{-3}$ & Ignore time \& frequency dependence. Artifacts due to time and frequency variations of the PB.\\
      
      \hline
      Top Right & \MTMFS + SI & Multi-term Imaging with Standard Gridding  & $1\times10^{-4}$ & $5\times10^{-4}$& Ignore time dependence. Absorb time-averaged frequency dependence in MT-MFS. Artifacts due to time-variability of the PB.\\
      
      \hline
      Lower Left & \MTMFS + \AWP\ &  Multi-term Imaging  with  \NBAWP\ gridding  & $4\times 10^{-5}$ & $8\times 10^{-4}$ & Account for time variability of PB, and absorb the resulting $PB^2$ frequency dependence in MT-MFS. Artifacts due to stronger spectral structure.\\
      
      \hline
      Lower Right & \MTMFS + \WBAWP\  & Multi-term Imaging with wide-band A-Projection gridding    &$3.5\times10^{-5}$&$2\times10^{-4}$  & Account for PB time- \& frequency-dependence in \WBAWP. Account for static sky-frequency dependence in MT-MFS. Minimal artifacts.\\
      \hline
    \end{tabular}
  }
  \label{FIG:RESULTS_SIM}
\end{figure*}


%

\begin{figure*}
  \centering
  \includegraphics[width=7in]{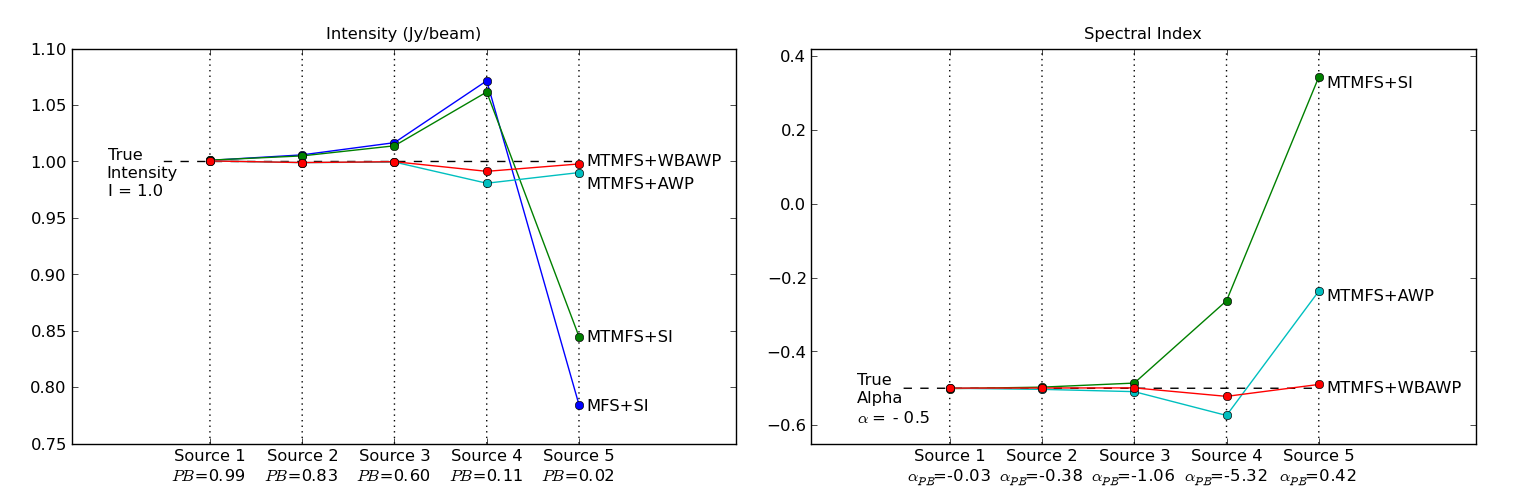}
  \caption{This figure compares the accuracy of the PB-corrected
    intensity (LEFT) and spectral-index (RIGHT) for the five simulated point sources,
    using the four methods whose results are shown in Figure
    \ref{FIG:RESULTS_SIM}. The labels ``{\tt AWP}'' and ``{\tt WBAWP}''
    are used for \AWP\ and \WBAWP\ in the figure.  The algorithms
    compared are MFS+SI, \MTMFS+SI, \MTMFS+\AWP\ and \MTMFS+\WBAWP.
    Spectral-indices are shown only for methods using \MTMFS, with
    post-deconvolution (average) spectral-index corrections done for
    the SI and \AWP\ runs.  Results for the five sources are shown from
    left to right with increasing distance from the pointing-center.
    The reference-PB gain and effective PB-spectral-index at the
    locations of the five sources are listed on the x-axis.  These
    plots show that outside the HPBW at the reference-frequency,
    methods that do not account for time-variable PB-spectra have
    considerably higher errors, and the combination of \MTMFS+\WBAWP\
    delivers accurate corrections even out in the sidelobe.  }
  \label{FIG:RESULTS_SIM_INTALPHA}
\end{figure*}

\subsection{Cube imaging + MFS deconvolution}
\label{Sec:cubegrid}
The simplest extension of NB-\AWP\ for multi-frequency synthesis is to
grid, Fourier transform and normalize each frequency channel (or sets
of channels) independently, and then produce a continuum image by an
image-domain accumulation, before the minor cycle.
%
%
When the sky spectrum is not flat, or when {\it flat-noise}
normalization is used per channel, a wide-band minor-cycle algorithm
such as \MTMFS\ can be applied to simultaneously solve for the sky  
intensity and spectrum. Taylor-weighted residual images are 
constructed as follows, before proceeding to the minor cycle.
\begin{equation}
  I^{R}_t = \sum_{\nu} \wnt  I^{R}_{\nu} = \sum_{\nu} \wnt \left\{ {\Pnu \cdot \left( \Inu^{psf} \star \left(\Pnu \cdot \Inu^{sky}\right) \right)} \over {\Pnu^2}   \right\}
\end{equation}
where $\wnt = \dnuno^t$ are weights that represent Taylor polynomial
basis functions \citep{MSMFS}.  The interpretation of the output Taylor
coefficients depends on the choice of normalization as follows.
\begin{enumerate}
\item With {\it flat-sky} normalization per channel, the output Taylor
  coefficients will represent $I^{sky}_{\nu}$.
\item For {\it flat-sky} normalization per channel, but a regularized
  {\it flat noise} minor cycle, the Taylor-weighted residual images
  can be multiplied by a $P_{ref}$ after the frequency summation and
  before deconvolution. The output Taylor coefficients will represent
  $P_{ref} I^{sky}_{\nu}$ and a post-deconvolution division 
by $P_{ref}$ will be required for the intensity image. No corrections are
needed for the spectral index map.
\item With {\it flat-noise} normalization per channel before
  Taylor-weighted averaging, the output Taylor coefficients will
  represent $P_{\nu} I^{sky}_{\nu}$, and a post-deconvolution
  polynomial division of the primary beam spectrum will be required
  to correct both the intensity and the spectral index.
\end{enumerate}

Qualitatively, the reverse transform with {\it flat-sky} normalization
per channel followed by frequency-averaging and a {\it flat-noise}
regularization by $P_{ref}$ before the minor cycle, is equivalent to
the use of \WBAWP\ during gridding.  It requires multiple image grid
planes (one per channel), FFTs and has repeated beam divisions and
multiplications that can increase numerical errors.  However, it is
naturally parallelizable, making it an attractive option for extremely
large data sets and imaging goals where inaccuracy in low gain regions
of the primary beam can be tolerated.

\subsubsection{MFS imaging + MFS deconvolution}
\label{Sec:MFSim.MFSdec}
To optimize on memory use and FFT costs (especially in a non-parallel
imaging run), MFS gridding can be done, where averages over baseline,
time and frequency are accumulated onto a single grid, followed by a
single FFT and normalization by an average primary beam (or its
square).
Here, attention must be paid to the consequences of averaging over
frequency before normalization.  The use of $\Adj{\A(\nu)}$ as the
  gridding convolution function (the NB-\AWP\ algorithm), introduces
  an additional frequency dependence $P_{\nu}$ that gets averaged over
  before it can be removed.  Once gridded, this extra frequency
  dependence is locked in, and can be accounted for only as an
  artificially steeper spectrum in the minor-cycle of the \MTMFS\
  wide-band imaging algorithm.

Multi-term Taylor-weighted residual images must be constructed as
follows,
\begin{equation}
  I^{R}_t =  { { \sum_{\nu} \wnt  \left\{ {\Pnu \cdot \left( \Inu^{psf} \star \left(\Pnu \cdot \Inu^{sky}\right) \right)} \right\} } \over {\sum_{\nu}  \Pnu^2} }
\end{equation}
and the output Taylor coefficients will depend on the choice of
normalization as follows
\begin{enumerate}
\item For {\it flat-sky} normalization, the model intensity represents
  $I^{sky}_{ref}$, but the spectrum represents the product of the sky
  spectrum and the square of the primary-beam spectrum.
\item 
  For {\it flat-noise} normalization, the model intensity represents $I^{sky}_{ref} P_{avg}$ and
  the spectrum represents the product of the sky spectrum and the square of the primary beam spectrum.
\end{enumerate}
In both cases, appropriate wide-band post-deconvolution corrections for the average primary beam
and two instances of the primary beam spectrum, must be applied.

Such corrections are inelegant, and are susceptible to numerical instabilities in low gain regions 
of the primary beams. Section~\ref{SEC:RESULTS_MTMFS} shows a comparison of some of these
methods with \WBAWP, for a simulation with source spectra that are not flat and therefore
require \MTMFS\ imaging and deconvolution. 

\subsection{Comparison of Hybrids with \WBAWP}
\label{SEC:RESULTS_MTMFS}

To test the algorithm described in sections~\ref{SEC:WBAWP_ALGO} and \ref{SEC:ATERM}
with non-flat source spectra, we used the sky brightness distribution as in 
Fig.~\ref{FIG:RESULTS_SIM_FLAT}, but assigned a spectral index of $\alpha = -0.5$ 
to all sources such that $I(\nu) \propto(\nu/\nu_o)^\alpha$.

Figure~\ref{FIG:RESULTS_SIM} shows deconvolved images produced with
and without time-dependent and frequency-dependent PB-corrections
during gridding, emphasizing the different types of error-patterns
that arise when one or more effects are ignored. 
An image formed from the hybrid method described in 
Sec.~\ref{Sec:MFSim.MFSdec} to absorb all frequency-dependence
into the minor cycle solver is also shown for comparison.
Figure~\ref{FIG:RESULTS_SIM_INTALPHA}
shows Stokes-I and spectral index values for these point-sources
after PB-correction, to illustrate the
accuracy to which different methods are able to recover the true-sky
spectral index at various locations in the PB.
The various methods tested and results obtained are described below.

\subsubsection{MFS~+~SI (Standard Imaging)}
\label{SEC:STANDARD_IMAGING}
The image in the top left panel of Fig.~\ref{FIG:RESULTS_SIM} is the
result of standard Cotton-Schwab Clean with MFS gridding using
prolate-spheroidal functions as gridding-convolution functions, and a
flat-spectrum assumption during the minor cycle. 
\begin{equation}
  I^{R} = \sum_{\nu} \left\{ { \left( \Inu^{psf} \star \left(\Pnu \cdot \Inu^{sky}\right) \right)}  \right\}
\end{equation}
Time and frequency variability of both the sky and the instrument
are ignored, and for a 66\% bandwidth, imaging artifacts
around all sources away from the pointing-center are dominated by
spectral-effects due to $P_{\nu}$ present in the data.
A post-deconvolution division by an average primary beam can recover the
true source intensity to within a few percent, out to the half-power point
of the PB, but errors increase with distance from the pointing center.

\subsubsection{\MTMFS~+~SI}
\label{SEC:METHOD_1}

The image in the top right panel of Fig.~\ref{FIG:RESULTS_SIM} is the
result of the \MTMFS\ algorithm in the minor cycle, with standard
gridding (prolate-spheroidal functions). The minor cycle solves for
the average intensity and spectrum of $I(\nu) P(\nu)$
using a 2-term Taylor-polynomial approximation.  
\begin{equation}
  I^{R}_t = \sum_{\nu}\wnt \left\{ { \left( \Inu^{psf} \star \left(\Pnu \cdot \Inu^{sky}\right) \right)}  \right\}
\end{equation}
Average PB-spectral effects are absorbed into the sky model, and the dominant remaining
error is due to the time-variability of the primary-beams.  A
post-deconvolution correction of the continuum intensity and spectral-index 
are accurate to within a few percent in intensity and $\pm$0.1 in spectral index
out to approximately the half-power point. Beyond this field-of-view, 
errors increase (to $\pm0.4$ or more in spectral index) primarily because 
a time-averaged primary-beam spectrum is not a good estimate in regions
of the image where $P_{\nu}$ changes by 100\% with time as the beams rotate on the sky.

\subsubsection{\MTMFS~+~\AWP}
\label{SEC:METHOD_2}

The image in the bottom left panel of Fig.~\ref{FIG:RESULTS_SIM} is
the result of \MTMFS\ in the minor cycle (2 terms), but with the
\NBAWP\ gridding as described in Sec.~\ref{Sec:MFSim.MFSdec}
with a {\it flat-noise} normalization before the minor cycle, followed
by a post-deconvolution correction of the intensity by $P_{ref}$ (average primary beam)
and the spectral index by twice that of the primary beam.
Artifacts due to frequency-independent time-variability (antenna
rotation) no-longer exist within the HPBW (PB gain of 0.5), but new
spectral artifacts appear away from the pointing-center (beginning
around the 10\% level).

These errors are partly due to the increased
non-linearity of the $P^2(\nu)$ spectrum away from the pointing center,
for which a two-term Taylor-polynomial approximation is insufficient. 
A run with 3 terms partially reduces this problem, indicating that errors in 
approximating the combined spectrum with a low-order polynomial dominates 
the errors, but higher order polynomials are inadvisable because of 
instability in low-SNR regions. 

Errors also arise from the high time variability of the PB-spectrum,
which is ignored because only time-averaged spectra are used for
spectral-correction.  A post-deconvolution correction of the
spectral-index map for $P^2(\nu)$ results in errors at the $\pm 0.3$ level
beyond the $\sim$50\% point.

\subsubsection{\MTMFS~+~\WBAWP}
\label{SEC:METHOD_3}

The image in the bottom right panel of Fig.~\ref{FIG:RESULTS_SIM} is
the result of \MTMFS\ in the minor cycle (two terms), and \WBAWP\
gridding (section~\ref{SEC:WBAWP_ALGO}).  Artifacts around all sources
are gone, and the dominant errors are numerical (at the floating-point
precision level).  
The spectral-index map produced by \MTMFS\ is accurate to within
0.01 in the main lobe, and 0.05 out in the sidelobe.  Such accuracies
allows the recovery of source-spectra further-out in the primary-beam
than previously possible.  The main difference between this method and
all others, is that time and frequency variability of the primary beam
has been corrected for in the data domain, before any averaging is done
to construct a continuum image to send to the minor cycle.  
The minor cycle sees a flat-noise normalization,
preserving the convolution-equation and allowing for deeper 'cleaning'
before triggering the next major cycle (i.e. faster convergence).

This method shows the lowest errors in Fig.~\ref{FIG:RESULTS_SIM_INTALPHA}
indicating that if the primary beam can be accurately modeled, its
time and frequency variability can be corrected for during
gridding, resulting in an accurate reconstruction in the minor cycle.

\subsection{Imaging results with VLA L-Band data}
\label{Sec:EVLAdata}
\begin{figure*}[ht!]
\centering
 \includegraphics[width=6.5in]{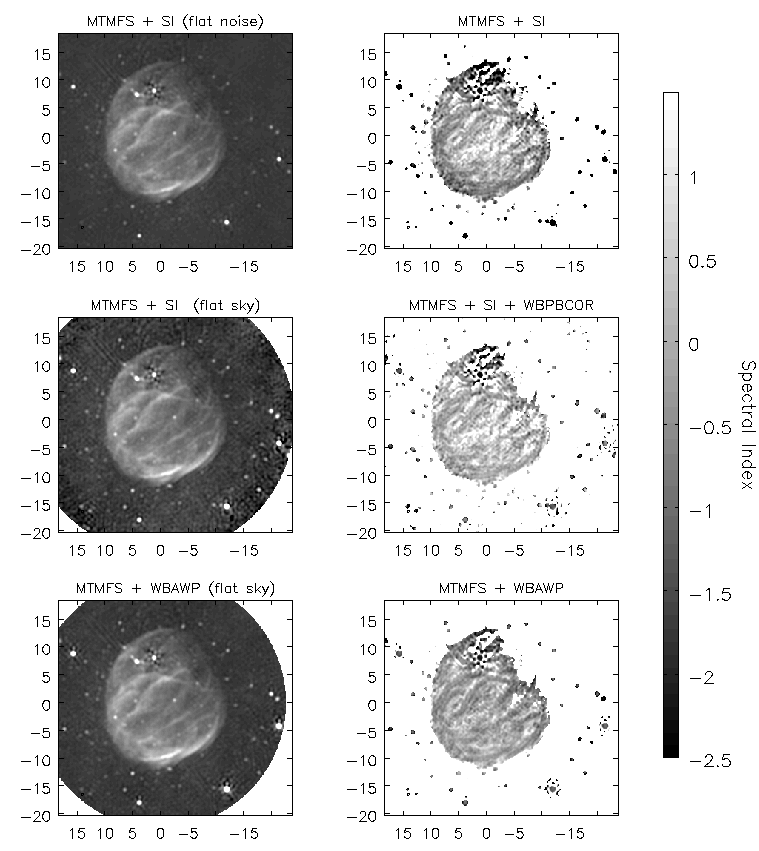}
 \caption{The left column shows wide-band continuum Stokes-I images of
   the Galactic SNR G55 imaged with the VLA centered at 1.5~GHz
   covering the frequency range $1.256 - 1.905$~GHz. The right column
   shows the spectral index maps.  The top and middle rows are the results
from MTMFS+SI imaging, without and with a post-deconvolution wideband
PB-correction respectively. 
   The artificially high spectral index around the edge of the SNR emission 
  and farther out, in the top row, is due to the PB. This is not present in the
middle row, but the spectral index map is still noisy.
The bottom row shows the flat-sky results from
MTMFS+WB\AWP\ with spectral indices closer to their expected values than
either of the other methods.  Details are discussed in Sec.~\ref{Sec:EVLAdata}.}
\label{FIG:RESULTS_G55}
\end{figure*}

Figure~\ref{FIG:RESULTS_G55} shows the continuum intensity and spectral
index distribution of the G55.7+3.4 Galactic supernova remnant (SNR),
using the VLA in L-band and D-configuration and 
made with and without the \WBAWP\ algorithm.  
The peak brightness is 6mJy, and with an off-source RMS of 11 $\mu$Jy,
this is a modest dynamic range. The peak brightness comes from a background
pulsar with a known spectral index of -2.3, and the brighter synchrotron-emission
filaments are at the 1mJy level.
The half-power beam width (HPBW) of the PB is 30 arcmin, and 
extended emission from the SNR fills the PB at the reference frequency.
The spurious spectral index at the HPBW due to the primary beam 
variation between 1-2 GHz is approximately -1.4. 

The left column of panels in Fig.~\ref{FIG:RESULTS_G55} 
shows continuum intensity, and the right column
shows corresponding spectral index maps.
\begin{enumerate}
\item The top row shows flat-noise results with MTMFS+SI, where A-Projection 
was not used, and primary-beam correction was not done. 
There is considerable artificial steepening (darkening on the plot)
of the observed source spectrum as distance from the pointing center increases. 
The spectral index of the bright background pulsar is -3.05.
\item The middle row shows flat-sky results from the same run as above,
where MTMFS+SI was followed by a post-deconvolution wideband PB correction
of both the intensity and the spectrum.
The spectral index map shows that this post-deconvolution correction has
restored the spectral indices of the outer part of the SNR as well as the background 
sources to more realistic values. 
The spectral index of the bright background pulsar is -2.61 (after correction for
an average estimated PB spectral index of -0.44 at the 0.8 gain level).
\item The bottom row shows flat-sky results from an MTMFS+WBAWP run
where the intensity image has been corrected for the average PB gain, 
but the spectral index image is just what came out of the 
imaging run. The noise properties of the intensity image are slightly better than the
middle row, and the spectral index map shows slightly more coherent and
less noisy structure across the SNR.
The spectral index of the bright background pulsar is -0.29, which is the closest
so far to the expected value. 
\end{enumerate}

\section{Conclusions}
\label{SEC:CONCLUSIONS}

In this paper we describe the wide-band \AWP\ (\WBAWP) algorithm,
which extends the narrow-band \AWP\ (\NBAWP) algorithm
\citep{AWProjection} to correct for the wide-band
effects of the PB {\it prior} to integration in time and frequency for
continuum imaging.  We demonstrate that the combined \WBAWP\ and
MT-MFS algorithm for simultaneous intensity and spectral index
mapping performs as expected.

Theoretical analysis in section~\ref{SEC:GENERALIZEDCALIB} draws
equivalence between standard antenna complex gain and bandpass
calibration and the \NBAWP\ and \WBAWP\ algorithms respectively and
show that the latter two algorithms are the direction-dependent
generalization of the former direction-independent algorithms. The
A-term of the \AWP\ algorithm represents the direction-dependent (DD)
complex antenna gain pattern in the data domain.  Since it is
direction-dependent, corrections for it fundamentally cannot be
decoupled from imaging and must be corrected for {\it during} imaging
\citep[see][]{IMAGING_THEORY_IEEE}.  The \AWP\ algorithm, which
corrects for the DD antenna gains during imaging, therefore can be
thought of as an algorithm for DD calibration.  Similarly, the \WBAWP\
algorithm which includes corrections for the frequency dependence of
the A-term can be thought of as the DD generalization of the standard
bandpass calibration algorithm.  We feel that making these connections
with simpler, intuitively better understood and widely used algorithms
in the community makes it easier to understand the newer more general
techniques.

We also analyzed hybrid schemes for wide-band imaging using the
\NBAWP\ and image-plane correction for the effects of PB.
Our conclusion is that while for non-parallel implementation, \WBAWP\ is
required for wide-band imaging, for implementations which may require
partitioning the data along time and/or frequency axis, hybrid
approaches are also sufficient.  However the use of \WBAWP\ in
implementations on parallel processing platforms allows the freedom to
tune the distribution of the data to suite the available hardware and
computing resources (e.g., this allows imaging smaller chunks of the
total bandwidth in parallel, even if these smaller chunks need
wide-band PB corrections.  Without \WBAWP, the data distribution is
restricted to be partitioned in frequency such that each chunk can
be imaged using \NBAWP).

Comparisons show that the \WBAWP\ plus \MTMFS\ enables simultaneous
intensity and spectral index imaging throughout the PB in
wide-field imaging. Moderate dynamic range imaging within the
half-power point of the PB is possible where all frequency dependence
in the image is absorbed in the solution of the \MTMFS\ algorithm.
Beyond this field-of-view (FoV), errors increase because a time-averaged
primary-beam spectrum is not a good estimate in regions of the image
where PB changes by 100\% with time as the beams rotate on the sky.
The FoV can be increased till $\sim 10\%$ point of the PB by combining
\MTMFS\ with \NBAWP.  The time-dependence of the PB is
accounted for via \AWP\ and its frequency dependence is absorbed in
\MTMFS.  Using larger number of Taylor-terms in \MTMFS\ improves the
imaging performance for simpler fields, but is inadvisable because of
instability in low-SNR regions.

Finally, we would like to note that while only the effects of the
antenna PB were included in the A-term used in this paper, other
antenna-based DD effect can also be easily included.  The effect of
non-isoplanatic ionospheric/atmospheric phases is comparable to the
effect of PB for wide-band wide-field imaging at low frequencies,
particularly with aperture-array antenna elements.  Similar effects
come from the irregularities in the water vapor content in the lower
atmosphere for imaging at high frequencies.  The effects due to
ionosphere/atmosphere and PB need to be corrected simultaneously,
often for wide-band data in full polarization.  It may be possible to
extend the \WBAWP\ algorithm presented here to include corrections for
ionospheric effects.  Work to test these extensions is underway and
will be reported in future publications.

\begin{acknowledgements}
  
  This work was done using the R\&D branch of the CASA code base.  We
  wish to thank the CASA Group for the underlying libraries.  We thank
  T. J. Cornwell for his very useful and detailed comments/suggestions
  as a referee. Part of this work was funded by the ALBiUS work
  package of the European Commission Radionet FP7 program.
\end{acknowledgements}

\bibliographystyle{apj}
\bibliography{WBAWP}

\end{document}